\documentclass{IOS-Book-Article}

\usepackage{mathptmx}
\usepackage{soul}\setuldepth{article}

\usepackage{algorithm}
\usepackage{algpseudocode}
\usepackage{amssymb}
\newtheorem{definition}{Definition}[section]
\newtheorem{example}{Example}[section]
\usepackage{booktabs}
\usepackage{listings}
\usepackage{makecell}

\DeclareMathAlphabet{\pazocal}{OMS}{zplm}{m}{n}
\SetMathAlphabet\pazocal{bold}{OMS}{zplm}{bx}{n}

\usepackage{makecell}
\usepackage{paralist}
\usepackage{subfig}
\usepackage{stmaryrd}
\usepackage{tikz}

\def\hb{\hbox to 11.5 cm{}}

\begin{document}

\newcommand{\Perez}{Pérez}
\newcommand{\leaf}{f\texttt{[}\alpha\texttt{]}}
\newcommand{\field}{f\texttt{[}\alpha\texttt{]}\texttt{\{}\phi\texttt{\}}}
\newcommand{\fieldp}{f\texttt{[}\alpha\texttt{]}\texttt{\{}\phi^\prime\texttt{\}}}
\newcommand{\lleaf}{\ell:f\texttt{[}\alpha\texttt{]}}
\newcommand{\lfield}{\ell:f\texttt{[}\alpha\texttt{]}\texttt{\{}\phi\texttt{\}}}
\newcommand{\fragment}{\texttt{on} \textrm{ } t\{\phi\}}
\newcommand{\fragmentp}{\texttt{on} \textrm{ } t\{\phi^\prime\}}
\newcommand{\fragmentdots}{\texttt{on} \textrm{ } t\{\dots\}}
\newcommand{\concat}{\phi\dots\phi}
\newcommand{\concatalt}{\phi_1\dots\phi_k}
\newcommand{\fieldconcat}{f\texttt{[}\alpha\texttt{]}\texttt{\{}\phi\dots\phi\texttt{\}}}
\newcommand{\keys}{\mathbf{kr}^{(f)}}
\newcommand{\keysa}[1]{\mathbf{kr}^{(\texttt{#1})}}
\newcommand{\keye}{\mathbf{ke}^{(f)}}
\newcommand{\keyea}[1]{\mathbf{ke}^{(\texttt{#1})}}
\newcommand{\keya}{\mathbf{ka}^{(\mathsf{a})}}
\newcommand{\keyaa}[2]{\mathbf{ka}^{(\texttt{#1},\texttt{#2})}}
\newcommand{\keyf}{\mathbf{kt}^{(t)}}
\newcommand{\evalsem}[3]{\llbracket #1 \rrbracket^{#2}_{#3}}

\pagestyle{headings}
\def\thepage{}
\begin{frontmatter}              

\title{Native Execution of GraphQL Queries over RDF Graphs Using Multi-way Joins}

\markboth{}{July 2023\hb}
\author[A]{\fnms{Nikolaos} \snm{Karalis}\orcid{0000-0002-0710-7180}%
\thanks{Corresponding Author: Nikolaos Karalis,   nkaralis@mail.uni-paderborn.de. This work has received funding from the European Union's Horizon 2020 research and innovation programme under the Marie Skłodowska-Curie grant agreement No 860801 and the European Union’s Horizon Europe research and innovation programme under grant agreement No 101070305. It has also been supported by the Ministry of Culture and Science of North Rhine-Westphalia (MKW NRW) within the project SAIL under the grant no NW21-059D.}},
\author[A]{\fnms{Alexander} \snm{Bigerl}\orcid{0000-0002-9617-1466}}
and
\author[A]{\fnms{Axel-Cyrille} \snm{Ngonga Ngomo}\orcid{0000-0001-7112-3516}}

\runningauthor{Karalis et al.}
\address[A]{DICE group, Department of Computer Science, Paderborn University}

\begin{abstract}\emph{Purpose:} The query language GraphQL has gained significant traction in recent years.
In particular, it has recently gained the attention of the semantic web and graph database communities and is now often used as a means to query knowledge graphs.
Most of the storage solutions that support GraphQL rely on a translation layer to map the said language to another query language that they support natively, for example SPARQL.
\emph{Methodology:} Our main innovation is a multi-way left-join algorithm inspired by worst-case optimal multi-way join algorithms. This novel algorithm enables the native execution of GraphQL queries over RDF knowledge graphs. 
We evaluate our approach in two settings using the LinGBM benchmark generator.
\emph{Findings: } The experimental results suggest that our solution outperforms the state-of-the-art graph storage solution for GraphQL with respect to both query runtimes and scalability. 
\emph{Value:} Our solution is implemented in  an open-sourced triple store, and is intended to advance the development of representation-agnostic storage solutions for knowledge graphs. 
\end{abstract}

\end{frontmatter}
\markboth{July 2023\hb}{July 2023\hb}

\section{Introduction}
Knowledge graphs serve as the data backbone of an increasing number of applications. Examples of such applications include search engines, recommendation systems, and question answering systems \cite{DBLP:journals/ftdb/WeikumDRS21, DBLP:journals/csur/HoganBCdMGKGNNN21}. Consequently, efficient storage and querying solutions for knowledge graphs are imperative. Many triple stores \cite{DBLP:conf/semweb/BigerlCBSSN20,DBLP:journals/vldb/ZouOCSHZ14,DBLP:journals/pvldb/YuanLWJZL13,DBLP:journals/pvldb/NeumannW08,url/blazegraph,url/virtuoso,url/graphdb,url/fuseki} and graph databases \cite{url/neo4j,url/janus} have hence been developed in recent decades. Used primarily by the semantic web community, triple stores process RDF knowledge graphs. A popular representation model among the graph database community is the property graph model \cite{DBLP:journals/csur/HoganBCdMGKGNNN21, DBLP:series/synthesis/2018Bonifati}.
While SPARQL is the designated query language for RDF,  multiple languages have been developed to query property graphs (e.g., Cypher \cite{DBLP:conf/sigmod/FrancisGGLLMPRS18} and Gremlin \cite{DBLP:conf/dbpl/Rodriguez15}).
Recently, GraphQL, a query language for APIs, has attracted the attention of both the graph database \cite{url/dgraph,DBLP:conf/grades/HartigH19} and the semantic web \cite{DBLP:conf/semweb/GleimHKD20,DBLP:journals/ijseke/Chaves-FragaPAC20, DBLP:conf/semweb/TaelmanSV18,url/hypergraphql} communities.
\par
The focus of the semantic web community regarding GraphQL has been on the development of query translation tools \cite{DBLP:conf/semweb/GleimHKD20, DBLP:conf/semweb/TaelmanSV18,url/hypergraphql}.
These tools translate GraphQL queries into SPARQL queries, which are then issued to a triple store.
A drawback of such solutions is that the results produced by triple stores need to be rewritten, since GraphQL dictates a strict response format.
This process adds a significant overhead to the execution of queries \cite[Table 3]{DBLP:conf/semweb/GleimHKD20}, especially in cases where the results are large.
To the best of our knowledge, there are no publicly available triple stores that treat GraphQL as a first-class citizen.
\par
While most constructs for processing basic graph patterns can be exploited in a straightforward manner for GraphQL processing, the formal semantics of GraphQL \cite{DBLP:conf/www/Hartig018} demand the use of left-join operations for the evaluation of GraphQL queries.
However, conventional two-way left-join operations are not suitable for the evaluation of GraphQL queries, as the results of such queries can be constructed incrementally \cite{DBLP:conf/www/Hartig018}.
We hence focus on presenting a \emph{novel multi-way left-join algorithm} inspired by worst-case optimal join algorithms \cite{DBLP:conf/pods/000118}, which can be used to enumerate GraphQL queries incrementally.
By implementing our approach into a state-of-the-art triple store, we provide the \emph{first publicly available triple store that treats GraphQL queries as first-class citizens}.
We carried out an extensive evaluation using a synthetic benchmark generator, namely LinGBM \cite{DBLP:conf/wise/ChengH22}, and the results suggest that our implementation is able to outperform a state-of-the-art graph storage solution providing GraphQL support, namely Neo4j.
\par
The rest of this paper is structured as follows. The preliminaries are provided in Section \ref{sec:prelim}. In Section \ref{sec:leftjoin}, we present our multi-way left-join algorithm and show how to natively evaluate GraphQL queries over RDF graphs. We evaluate our approach in Section \ref{sec:evaluation}. We discuss related works in Section \ref{sec:relwork} and conclude in Section \ref{sec:conclusion}.

\section{Preliminaries}\label{sec:prelim}
Here, we introduce the concepts and the semantics of GraphQL that we use throughout this work along with their formal definitions as per \cite{DBLP:conf/www/Hartig018}.
We also briefly introduce worst-case optimal multi-way join algorithms, which have inspired our proposed algorithm.
\subsection{GraphQL}\label{sec:graphql}
GraphQL is a query language that was designed to simplify communication between clients and application servers.
One of the main characteristics of GraphQL is that it is strongly typed.
GraphQL services---i.e., servers and data sources whose data can be accessed and modified via GraphQL operations---expose a GraphQL schema to their clients by which incoming requests must abide.
This schema defines a type system that describes the structure of the underlying data of the GraphQL service and the operations the service supports.
Another important aspect of GraphQL is the hierarchical structure of its operations and responses.
GraphQL operations form a tree structure that specifies the traversal on top of the underlying graph and the information that should be extracted from the nodes at each step of the traversal.
In turn, the responses should follow the hierarchy defined by their respective operation.
The syntax and capabilities of GraphQL, as well as the responsibilities of GraphQL services, are detailed in the language's official specification \cite{url/graphql}.
Even though the specification describes how services should handle the requests they receive, it does not provide a formal specification of the semantics of the language.
Consequently, studying the expressiveness and complexity of the language remained a challenge.
To tackle the lack of formal semantics and the consequences thereof, Hartig and \Perez{} \cite{DBLP:conf/www/Hartig018} provide formal semantics for GraphQL queries that consist of \emph{fields}, \emph{field aliases} and \emph{inline fragments}.
The semantics rely on the formal definition of GraphQL schemata and graphs as well as the formalized syntax of GraphQL queries. Here, we reintroduce the definitions presented in \cite{DBLP:conf/www/Hartig018}.
\par
The formal definitions presented below rely on the following sets.
Let \textit{Fields} be an infinite set of field names and $F \subset \textit{Fields}$ a finite subset of $\textit{Fields}$.
Let $A$ and $T$ be finite sets of argument names and type names, respectively, where $T$ is the disjoint union of $O_T$ (object type names), $I_T$ (interface type names), $U_T$ (union type names) and \textit{Scalars} (scalar type names).
Last, let $L_T = \{ [t] \mid t \in T \}$ be the set of list types constructed from T and \textit{Vals} be a set of scalar values.
\begin{definition}[GraphQL schema  \cite{DBLP:conf/www/Hartig018}]
    A GraphQL schema $\pazocal{S}$ over $(F, A, T)$ is composed of the following components:
    \begin{itemize}
        \item $\textit{fields}_{\pazocal{S}} : (O_T \cup I_T) \rightarrow 2^F$ that assigns a set of fields to every object type and every interface type,
        \item $\textit{args}_{\pazocal{S}} : F \rightarrow 2^A$ that assigns a set of arguments to every field,
        \item $\textit{type}_{\pazocal{S}} : F \cup A \rightarrow T \cup L_T$ that assigns a type or a list type to every field and argument, where arguments are assigned scalar types,
        \item $\textit{union}_{\pazocal{S}} :  U_T \rightarrow 2^{O_T}$ that assigns a nonempty set of object types to every union type,
        \item $\textit{impl}_{\pazocal{S}}: I_T \rightarrow 2^{O_T}$ that assigns a set of object types to every interface,
        \item $\textit{root}_{\pazocal{S}} \in O_T$ that represents the \emph{query} root type.
    \end{itemize}
\label{def:schema}
\end{definition}
\begin{example}\label{ex:graphqlschema}
    Consider the following GraphQL schema $\pazocal{S}$
    \vspace{-1\baselineskip}
    \begin{center}

\begin{lstlisting}[basicstyle=\footnotesize\ttfamily]
    interface Entity {         type Company impl Entity {
      id:String                  id:String
      email:String               name:String
    }                            email:String
    type Person impl Entity {    employees:[Person]
      id:String                }
      fname:String             type Query {
      lname:String               people(lname:String):[Person]
      email:String               companies:[Company]
      age:Int                  }
    }                          schema { query:Query }
\end{lstlisting}%
\end{center}
Let $F = \{ $ \texttt{people, companies, employees, fname, age, id, lname, email, name} $\}$, $A = \{$ \texttt{lname} $\}$, $O_T = \{ $ \texttt{Query, Company, Person} $\}$, $I_T = \{$ \texttt{Entity} $\}$, $U_T = \{ \}$, and $\textit{Scalars} = \{$  \texttt{String, Int} $\}$. 
The GraphQL schema $\pazocal{S}$ is formalized over $(F,A,T)$ 
as follows (we omit several assignments for brevity):
\begin{center}
    $\begin{array}{l}
        \textit{args}_\pazocal{S}(\texttt{people}) = \{\texttt{lname}\}\textrm{, } \textit{fields}_\pazocal{S}(\texttt{Entity}) = \{\texttt{id, email}\}\textrm{, } \textit{type}_\pazocal{S}(\texttt{id}) = \texttt{String}\textrm{, } \\
        \textit{fields}_\pazocal{S}(\texttt{Person}) = \{\texttt{id, fname, lname, email, age}\}\textrm{, } \textit{root}_\pazocal{S} = \texttt{Query}.
    \end{array}$
\end{center}
\end{example}
In practice, GraphQL services are implemented on top of data sources that adopt different data models (e.g., relational databases and graph databases).
To provide the semantics of GraphQL queries in a unified manner, Hartig and \Perez{} \cite{DBLP:conf/www/Hartig018} introduced the notion of GraphQL graphs. 
GraphQL graphs are logical constructs that abstract the underlying data sources of GraphQL services.
\begin{definition}[GraphQL graph  \cite{DBLP:conf/www/Hartig018}]
A GraphQL graph over $(F, A ,T)$ is a tuple $G = (N, E, \tau, \lambda, \varrho)$ with the following elements:
\begin{itemize}
    \item $N$ is a set of nodes,
    \item $E$ is a set of edges of the form $(u, f[a], v)$, where $u, v \in N$, $f \in F$, and $a$ is a partial mapping from $A$ to $\textit{Vals}$,
    \item $\tau : N \rightarrow O_T$ is a function that assigns a type to every node,
    \item $\lambda$ is a partial function that assigns a scalar value $\nu \in \textit{Vals}$ or a sequence $[\nu_1 \dots \nu_n]$ of scalar values ($\nu_i \in \textit{Vals}$) to some pairs of the form $(u, f[a])$ where $u \in N$, $f \in F$ and $a$ is a partial function from $A$ to \textit{Vals},
    \item $\varrho \in N$ is a distinguished node called the root node.
\end{itemize}
\label{def:graphqlgraph}
\end{definition}
\begin{definition}[GraphQL query \cite{DBLP:conf/www/Hartig018}]
    A GraphQL query over $(F, A, T)$ is an expression $\phi$ constructed from the following grammar where \texttt{[,],\{,\},:} and \texttt{on} are terminal symbols, $t \in O_T \cup I_T \cup U_T,$ $f \in F$, $\ell \in \textit{Fields}$, and $\alpha$ is a  partial mapping from $A$ to \textit{Vals}:
    \begin{itemize}
        \centering
        \item [] $\phi ::= \leaf \mid \lleaf \mid \fragment \mid \field \mid \lfield \mid \concat$ .
    \end{itemize}
\label{def:graphqlquery}
\end{definition}
\begin{example}\label{ex:graphqlquery}
    Examples of GraphQL queries conforming to the GraphQL schema $\pazocal{S}$ of Example \ref{ex:graphqlschema} are the following:
    \begin{center}
    $\begin{array}{l}
        \phi_1 = \texttt{people(lname: "Doe") \{ fname email \} } \textrm{ and} \\
        \phi_2 = \texttt{companies \{ name employees \{ id lname \} \} }.
    \end{array}$
    \end{center}
    Both queries demonstrate the hierarchical structure of GraphQL queries.
    For example, $\phi_2$ accesses fields in the first level that belong to the object type \texttt{Company}. In the second level, it accesses fields of the object type \texttt{Person}, as $\textit{type}_\pazocal{S}(\texttt{employees}) = \texttt{[Person]}$. 
\end{example}
GraphQL queries of particular interest are those that are \emph{non-redundant} and in \emph{ground-typed} normal form.
According to \cite[Theorem 3.8]{DBLP:conf/www/Hartig018}, every GraphQL query can be transformed into an equivalent query that is non-redundant and in ground-typed normal form.
An important characteristic of such queries is that their response can be constructed without being subjected to \emph{field collection} \cite[Section 6.3.2]{url/graphql}.
This allows non-redundant GraphQL queries in ground-typed normal form to be evaluated in time linear to the size of their response \cite[Corollary 4.3]{DBLP:conf/www/Hartig018}.
\begin{figure}[t]
    \begin{minipage}{\textwidth}
        \raggedleft
        \input{graphqlsemantics}
    \end{minipage}
    \label{fig:graphqlsem}
\end{figure}
\begin{definition}[GraphQL semantics \cite{DBLP:conf/www/Hartig018}]\label{def:graphqlsemantics}
Let $G = (N, E, \tau, \lambda, \varrho)$ be a GraphQL graph and $\phi$ a non-redundant GraphQL query in ground-typed normal form, both conforming to a GraphQL schema $\pazocal{S}$ over $(F,A,T)$. The evaluation of $\phi$ over $G$ from node $u \in N$, denoted by $\evalsem{\phi}{u}{G}$, is captured by Equation \ref{eq:g_semantics}.\footnote{The expressions $\lfield{}$ and $\lleaf{}$ are evaluated by replacing $f$ with $\ell$ in the first two rules' results.}
The evaluation of $\phi$ over $G$, denoted by $\evalsem{\phi}{}{G}$, is simply $\evalsem{\phi}{\varrho}{G}$.
\end{definition}
In this work, we assume that $A \subset F$.
More specifically, we restrict the set of arguments of a field $f \in F$ to be the set of scalar fields of its type, i.e., $\textit{args}_{\pazocal{S}}(f) \subseteq \{f^\prime \mid f^\prime \in \textit{fields}_{\pazocal{S}}(\textit{type}_{\pazocal{S}}(f)), \textit{type}_{\pazocal{S}}(f^\prime) \in \textit{Scalars}\}$.
Hence, leaf fields are not assigned any arguments, and the expressions $\leaf$ and $\lleaf{}$ can be written as $f$ and $\ell:f$, respectively \cite{DBLP:conf/www/Hartig018}.
In \cite{DBLP:conf/www/Hartig018}, the sets $F$ and $A$ are assumed to be disjoint; however, our assumption is in accordance with the GraphQL specification and does not affect the provided semantics.

\subsection{Worst-case Optimal Multi-way Join Algorithms}\label{sec:wcoj}
Worst-case optimal multi-way algorithms \cite{DBLP:journals/jacm/NgoPRR18} have recently gained a lot of attention (e.g., \cite{DBLP:conf/semweb/BigerlCBSSN20,DBLP:conf/semweb/HoganRRS19,DBLP:conf/sigmod/ArroyueloHNRRS21,DBLP:journals/pvldb/FreitagBSKN20}) and  have demonstrated high performance in evaluating graph pattern queries \cite{DBLP:conf/semweb/HoganRRS19,DBLP:journals/pvldb/FreitagBSKN20,DBLP:conf/icde/KalinskyHMEK22}.
Such algorithms satisfy the AGM bound \cite{DBLP:conf/focs/AtseriasGM08} and their runtime matches the worst-case size of the result of the input query \cite{DBLP:conf/pods/000118,DBLP:conf/semweb/HoganRRS19}. Pair-wise join algorithm carry out join operations on two join operands at a time.
Instead, worst-case optimal multi-way algorithms (e.g., Leapfrog Triejoin \cite{DBLP:conf/icdt/Veldhuizen14}) are recursive and evaluate input queries on a per variable basis.
This evaluation method does not store any intermediate results and allows for solution mappings to be  directly written to the result.

\section{Evaluation of GraphQL Queries over RDF Graphs}\label{sec:leftjoin}
In this section, we introduce the multi-way left-join algorithm that we developed for the native execution of GraphQL queries over RDF graphs.
Motivated by recent results on the evaluation of basic graph pattern queries presented in \cite{DBLP:conf/semweb/BigerlCBSSN20,DBLP:conf/semweb/HoganRRS19}, the proposed left-join algorithm is inspired by worst-case optimal multi-way join algorithms and evaluates queries on a per variable basis.
However, unlike join operations, the reordering of left-join operations is not allowed.
Hence, we have to pay attention to the order in which the variables of a query are evaluated.
In addition, left-join operations might produce partial solutions (i.e., solutions with \texttt{null} values in the context of GraphQL). 
To respect the order of operations during the evaluation of a query and to ensure that partial solutions will not be discarded, we additionally introduce the \emph{operand dependency graph}.
Before introducing the operand dependency graph and the proposed multi-way left-join algorithm, we define first the process of generating the query operands of GraphQL queries.
\subsection{GraphQL Query Operands} \label{sec:gqlrdf}
In the case of SPARQL, there are multiple features of the language that generate query operands, with the most common being the triple pattern.
In the case of GraphQL, an operand needs to be generated for each \emph{field}, \emph{argument} and \emph{inline fragment} of a query.
Here, for simplicity, we use a notation that resembles SPARQL's triple patterns and present how to generate the operands of GraphQL queries.
Note that we do not actually translate GraphQL queries to SPARQL queries.
Potential implementations are free to use any means available (e.g., indices) for generating these operands.
For the generation of GraphQL query operands, we must also map the types and fields of the provided GraphQL schema to RDF terms. 
Our implementation computes this mapping using a GraphQL \emph{directive} \cite[Section 3.13]{url/graphql}.
In the following, we omit this mapping for brevity.
\par
In a GraphQL query, we distinguish three types of fields: \begin{inparaenum}[i)] \item \emph{root} fields, \item \emph{inner} fields, and \item \emph{leaf} fields\end{inparaenum}.
The root field of a query is the starting point of the traversal.
Its corresponding operand should only contain the entities of the underlying RDF graph that are instances of its type.
The pattern $\langle ?var, \texttt{rdf:type}, \textit{type}_\pazocal{S}(f) \rangle$ is used to extract these instances, with $?var$ being a variable that will be assigned the extracted instances.
Inner and leaf fields represent edges between a source and a target vertex in the graph and their operands are created using patterns of the form $\langle ?var_1, f, ?var_2 \rangle$.
Ultimately, $?var_1$ will be assigned the source vertices of the edge, whereas $?var_2$ will be assigned the target vertices.
In the case of inner and leaf fields, we need to also consider the type of the target vertices.
More specifically, in RDF, the objects of properties can vary in type, whereas, in GraphQL, the target vertices of fields are of specific type.
To restrict the type of target vertices, an additional operand is generated using the pattern of root fields presented above.
In practice, this additional operand can be omitted, if the schema allows it (e.g., via a directive).
Provided an expression $\field{}$, the operand of an argument-value pair $\mathsf{a} = (f^\prime, v) \in \alpha$ is created by $\langle ?var, f^\prime, v \rangle$.
Last, the operand of an inline fragment $\fragment{}$, whose sub-expression $\phi$ is executed only if the parent field is an instance of the type $t$, is created by $\langle ?var, \texttt{rdf:type}, t \rangle$.
The operands of the aliased fields $\lfield{}$ and $\lleaf{}$ are generated using the patterns of $\field{}$ and $\leaf{}$, respectively.
\par
Two query operands participate in a (left-)join operation, if they share a variable.
For assigning variables to operands, we take advantage of the hierarchical structure of GraphQL.
More specifically, the target vertices of a field and the source vertices of its nested fields, share the same variable.
The operands of inline fragments are also assigned the variable of the target vertices of their parent fields.
In the case of arguments, their operands are assigned the variable that is already assigned to the operand of their field.
Example \ref{ex:qoperands} demonstrates the operand generation process of GraphQL queries.
\begin{example}\label{ex:qoperands}
Consider the queries of Example \ref{ex:graphqlquery}.
The operands of $\phi_1$ are generated by the patterns: \begin{inparaenum}[1)] \item $\langle ?x, \texttt{rdf:type}, \texttt{Person} \rangle$, \item $\langle ?x, \texttt{lname}, \texttt{"Doe"} \rangle$, \item $\langle ?x, \texttt{fname}, ?y \rangle$, and \item $\langle ?x, \texttt{email}, ?z \rangle$ \end{inparaenum}.
Note that the operands of the root field and its argument share the same variable.
Consequently, vertices representing people whose last name is not ``Doe" will be discarded.
The inner fields are associated with the root field through the variable $?x$.
Also note that the target vertices of the inner fields are assigned different variables.
The operands of $\phi_2$ are created in a similar manner and their corresponding patterns are: \begin{inparaenum}[1)] \item $\langle ?x, \texttt{rdf:type}, \texttt{Company} \rangle$, \item $\langle ?x, \texttt{name}, ?y \rangle$, \item $\langle ?x, \texttt{employees}, ?z \rangle$, \item $\langle ?z, \texttt{rdf:type}, \texttt{Person} \rangle$, \item $\langle ?z, \texttt{lname}, ?w \rangle$, and \item $\langle ?z, \texttt{id}, ?v \rangle$. \end{inparaenum} The inner fields \texttt{name} and \texttt{employees} are associated with the root field \texttt{companies} through the variable $?x$, whereas the operands of the leaf fields \texttt{id} and \texttt{lname} are associated with the operand of their parent field, namely \texttt{employees}, through $?z$.
Last, note the additional operand that is generated for the field \texttt{employees}. Its goal is to discard vertices that are not of type \texttt{Person}. We assume that type filtering is not required for scalar types for brevity.
\end{example}

\subsection{Operand Dependency Graph}\label{sec:odg}
The operand dependency graph is inspired by \emph{pattern trees} \cite{DBLP:conf/pods/LetelierPPS12} and captures the dependencies between the operands of a query.
If an operand is not successfully resolved during the query evaluation, its dependent operands should not be evaluated.
For example, provided a GraphQL query $\field{}$, the operands of $\phi$ should not be considered if the operands of $\leaf$ do not produce any results.
However, if the operands of $\phi$ do not produce any results, the results generated by $\leaf{}$ should not be discarded.
The operand dependency graph is formally defined as follows.
\begin{definition}[Operand dependency graph]\label{def:odf}
Let $O$ be a list of query operands and $\Sigma$ an alphabet.
Furthermore, let $\mathbb{I}_n = \{ i \in \mathbb{N} \mid 1 \leq i \leq n \}$.
An operand dependency graph is a directed vertex-edge-labelled graph $G = (V, E)$, where $V = \mathbb{I}_{|O|}$ and $E \subseteq V \times \Sigma \times V$.
An operand $v \in V$ depends on operand $u \in V$, if and only if $\exists e \in E$ such that e = $(u, \sigma, v)$  and $\sigma \in \Sigma$.
\end{definition}
As per Definition \ref{def:odf}, the vertices of an operand dependency graph correspond to the operands of its respective query.
The variables appearing in query operands are assigned unique labels stemming from $\Sigma$ and are used to label the vertices and edges of the dependency graph.
The vertices of the dependency graph are assigned the labels of their respective operands' variables.
Two operands are connected via an edge only if they share a variable.
The label of an edge is determined by the label shared by its incident vertices.
\par
For the construction of the operand dependency graph, we take advantage of the hierarchical structure of GraphQL queries.
Provided an expression $\field{}$ ($\lfield{}$), the operands of $\leaf{}$ comprise a strong component in the dependency graph, as they all depend on each other.
This means that any vertex $v$ of $\leaf{}$ is reachable from any other vertex $u$ of $\leaf{}$, with $v \ne u$, provided that $\leaf{}$ generates multiple operands.
As the operands of $\phi$ depend on the operands of $\leaf{}$, the vertices of $\leaf{}$ are not reachable from the vertices of $\phi$.
In the case of $\fragment{}$ expressions, the operands of $\phi$ depend on the operand of $t$.
This means that the vertex of $t$ and the vertices of $\phi$ are connected with edges, whose source is the vertex corresponding to $t$.
In the case of $\concatalt{}$ expressions, there are not any edges between the vertices of any $\phi_i$ and $\phi_j$, with $ 1 \leq i,j \leq k$ and $ i \neq j $, as the evaluation of $\phi_i$ does not affect the evaluation of $\phi_j$ (Definition \ref{def:graphqlsemantics}).
Operands that depend on each other participate in join operations, whereas unidirectional edges denote left-join operations.
Last, in the multi-way left-join algorithm, which is presented below, we make use of the root node of the directed acyclic graph connecting the strongly connected components of an operand dependency graph. 
Herein, we refer to this node as the \emph{independent strong component} of the dependency graph.
\begin{example}
The operand dependency graphs corresponding to the GraphQL queries Example \ref{ex:graphqlquery} and their respective operands (Example \ref{ex:qoperands}) are as follows.
\begin{center}
\begin{tikzpicture}[scale=0.8]
    \node[shape=circle,draw=black,scale=0.9,label=above:{$x$}] (1) at (1,0.5) {$1$};
    \node[shape=circle,draw=black,scale=0.9,label=above:{$x$}] (2) at (2.5,0.5) {$2$};
    \node[shape=circle,draw=black,scale=0.9,label=above:{$x,y$}] (3) at (4,1)  {$3$};
    \node[shape=circle,draw=black,scale=0.9,label=below:{$x,z$}] (4) at (4,0) {$4$};
    \path [->] (1) edge[bend left] node[above] {$x$} (2);
    \path [->] (2) edge[bend left] node[below] {$x$} (1);
    \path [->]  (2) edge node[above] {$x$} (3);
    \path [->]  (2) edge node[below] {$x$} (4);e
    
    \node[shape=circle,draw=black,scale=0.9,label=above:{$x$}] (10) at (8,0.5) {$1$};
    \node[shape=circle,draw=black,scale=0.9,label=above:{$x,y$}] (11) at (10,1)  {$2$};
    \node[shape=circle,draw=black,scale=0.9,label=below:{$x,z$}] (12) at (10,0) {$3$};
    \node[shape=circle,draw=black,scale=0.9,label=below:{$z$}] (13) at (12,0) {$4$};
    \node[shape=circle,draw=black,scale=0.9,label=above:{$z,w$}] (14) at (14,1) {$5$};
    \node[shape=circle,draw=black,scale=0.9,label=below:{$z,v$}] (15) at (14,0) {$6$};
    \path [->]  (10) edge node[above] {$x$} (11);
    \path [->]  (10) edge node[below] {$x$} (12);
    \path [->]  (12) edge[bend left] node[above] {$z$} (13);
    \path [->]  (13) edge[bend left] node[below] {$z$} (12);
    \path [->]  (13) edge node[above] {$z$} (14);
    \path [->]  (13) edge node[below] {$z$} (15);
\end{tikzpicture}
\end{center}
In the operand dependency graph of $\phi_1$ (left-hand side), there are not any edges connecting the vertices of operands $3$ and $4$, since the evaluation of $3$ does not affect the evaluation of $4$, and vice versa.
The independent strong component of the dependency graph consists of the operands $1$ and $2$.
In the operand dependency graph of $\phi_2$ (right-hand side), operands $3$ and $4$ depend on each other.
Both operands are generated by the inner field \texttt{employees}, with operand $4$ being responsible for removing any RDF terms assigned to $?z$ that are not of type \texttt{Person}.
In this case, the independent strong component consists of a single vertex, namely the vertex corresponding to operand $1$.
\end{example}
\subsection{Multi-way Left-Join Algorithm}
Here, we present our multi-way left-join algorithm (Algorithm \ref{alg:mwlj}) for the evaluation of GraphQL queries over RDF graphs.
The key characteristics of our approach are the following. 
First, it evaluates join and left-join operations on a variable simultaneously. Second, it uses the operand dependency graph to eliminate the transitively dependent operands of an empty operand (i.e., an operand that is not successfully resolved), thus avoiding unnecessary operations.
\par
The function \texttt{MWLJ} (lines. 1--5) takes as input a GraphQL query and is responsible for generating the operands of the query (line 2) and their dependency graph (line 3).
For simplicity, we assume throughout the algorithm that the operands are stored within the vertices of the graph.
This function is also responsible for initializing the solution mapping, which stores the bindings of all variables of the query, as its domain is equal to the set of labels appearing in the query's operand dependency graph.
Recall that each variable is assigned a unique label (Section \ref{sec:odg}).
After initializing the solution mapping, \texttt{MWLJ} calls the recursive function \texttt{MWLJ\_REC} (line 5), which takes the operands dependency graph $G$ and the solution mapping $X$ as inputs.
\begin{algorithm}[t]
    \footnotesize
    \algtext*{EndIf}
    \algtext*{EndFor}
    \algtext*{EndFunction}
    \caption{Multi-way Left-join Algorithm}
    \label{alg:mwlj}
    \begin{algorithmic}[1]
    \Function{MWLJ}{$Q$} \Comment{$Q$: Input GraphQL query}
        \State $O$ $\leftarrow$ generate the operands of $Q$
        \State $G$ $\leftarrow$ create the operand dependency graph of $Q$ using $O$
        \State $X$ $\leftarrow$ initialize solution mapping with domain equal to the set of labels appearing in $G$
        \State \Call{MWLJ\_REC}{$G$, $X$}
    \EndFunction
    \Function{MWLJ\_REC}{$G$, $X$} \Comment{$G$: operand dependency graph, $X$: solution mapping}
        \If{$G$ is disconnected}\Comment{Evaluation of $\concatalt{}$ expressions}
            \ForAll{connected components $G_i$ of $G$} \Comment{Each $G_i$ corresponds to a $\phi_i$, $1 \le i \le k$}
                \State \Call{MWLJ\_REC}{$G_i$, $X$}
            \EndFor
        \ElsIf{$G$ is not \emph{strongly connected}} \Comment{Left-join operation}
            \State $U$ $\leftarrow$ the set of labels appearing in the independent strong component of $G$
            \State $x$ $\leftarrow$ select a label from $U$
            \ForAll{values $\chi$ of $x$}
                \State resolve $x$ in all operands using $\chi$ \Comment{Carries out join and left joins simultaneously}
                \State $G^\prime$ $\leftarrow$ prune vertices of empty operands and their transitively dependent vertices from $G$
                \If{$G^\prime$ is empty}
                    \State \textbf{continue} \Comment{All operands are pruned (unsuccessful join); continue with the next $\chi$}
                \EndIf
                \State update the value of $x$ in $X$ with $\chi$ \Comment Join operations were successful
                \State remove $x$ from $G^\prime$; remove vertices without any labels from $G^\prime$
                \State \Call{MWLJ\_REC}{$G^\prime$, $X$}
            \EndFor
        \Else \Comment{$G$ is strongly connected (no left-join operations)}
            \State \Call{MWJ}{$G$, $X$} \Comment{Carry out multi-way join (no more left joins after this point)}
        \EndIf
    \EndFunction
    \end{algorithmic}
\end{algorithm}
\par
The function \texttt{MWLJ\_REC} (lines 6--22) is responsible for carrying out the join and left-join operations and generating the solutions of the query.
In case the provided dependency graph is disconnected, \texttt{MWLJ\_REC} is called for each connected component of the graph (lines 7--9).
Disconnected dependency graphs correspond to $\concatalt{}$ expressions, as there are no dependencies between any $\phi_i$ and $\phi_j$, with $ 1 \leq i,j \leq k$ and $ i \neq j$ (Section \ref{sec:odg}).
If the provided graph is not \emph{strongly} connected, there are left-join operations that need to be carried out (lines 10--20). 
To respect the order of left-join operations, the algorithm focuses on the set of labels (i.e., variables) $U$ that are found in the independent strong component of the dependency graph (line 11).
For the GraphQL queries that we consider in this work, the set $U$ contains only a single label.
This will not be the case once we take GraphQL's \emph{input object types} \cite[Section 3.10]{url/graphql} into consideration.\footnote{Note that also the formal definitions of GraphQL in \cite{DBLP:conf/www/Hartig018} do not consider input object types.}
After selecting a label $x$ from $U$, the algorithm iterates over all possible values of $x$ and carries out all join and left-join operations on $x$ (line 14).
The algorithm proceeds by removing any operands that were not successfully resolved along with their transitively dependent operands, which can be found by traversing the dependency graph (line 15).
If the resulting graph $G^\prime$ ends up being empty, a join operation was not successful and the algorithm continues with the next value of $x$ (lines 16--17).
If $G^\prime$ is not empty, the solution mapping $X$ is updated with the current value of $x$, which is removed from $G^\prime$ along with any fully resolved operands, and the algorithm proceeds with the next recursive step (lines 18--20).
In case the provided graph $G$ is \emph{strongly} connected, the algorithm proceeds with a multi-way join algorithm, as there are no left-join operations left to be carried out.
The active solution mapping $X$ will be ultimately projected once the remaining join operations are carried out by the multi-way join algorithm.
\begin{example}\label{ex:mwlj}
Consider the query $\phi_1$ of Example \ref{ex:graphqlquery}. Provided the example RDF graph
\begin{verbatim}
<p1> rdf:type <Person>; <lname> "Doe"; <fname> "Jon"; <email> "e1".
<p2> rdf:type <Person>; <lname> "Doe"; <fname> "Jan".
\end{verbatim}
the proposed algorithm will produce three solutions: \texttt{\{x:p1, y:"Jon"\}}, \texttt{\{x:p1, z:"e1"\}}, and \texttt{\{x:p2, y:"Jan"\}}.
The algorithm selects first the label corresponding to the variable \texttt{x}, which is assigned the identifiers of people in the graph.
For the value \texttt{p1} of \texttt{x}, the algorithm generates two solutions.
The first one provides the first name (\texttt{fname}) of \texttt{p1}, which is assigned to \texttt{y}, whereas the second one provides its email, which is assigned to \texttt{z}.
For the value \texttt{p2}, the algorithm generates only one solution, as \texttt{p2} does not have an email in the example graph.
Note that after selecting \texttt{$x$} and removing it from the operand dependency graph, the resulting dependency graph is disconnected.
Variables that do not appear in a solution mapping are unbound in that particular mapping.
\end{example}
\par
Regarding the enumeration of GraphQL queries, in \cite{DBLP:conf/www/Hartig018}, the authors study the enumeration problem for GraphQL queries that are non-redundant and in ground-typed normal form. 
Recall that such queries can be computed in time linear to the size of their response (Section \ref{sec:graphql}).
Each solution mapping generated by our algorithm captures a unique path of the response corresponding to the provided query.
As our left-join algorithm computes a solution mapping entirely, we are able to directly construct the path that corresponds to a particular solution mapping, once it is evaluated. 
In addition, due to the recursive nature of our algorithm, the solution mappings of the sub-trees of a particular node of a GraphQL response share common values (Example \ref{ex:mwlj}).
Hence, we are able to avoid visiting the nodes of a response multiple times.
\subsection{Implementation}\label{sec:impl}
We have implemented the proposed algorithm within the tensor-based triple store Tentris \cite{DBLP:conf/semweb/BigerlCBSSN20}.
Tentris achieves state-of-the-art performance in the evaluation of basic graph patterns, which are evaluated by a worst-case optimal multi-way join algorithm \cite{DBLP:conf/semweb/BigerlCBSSN20,DBLP:conf/semweb/BigerlCBSN22}.
Our implementation, namely TentrisGQL, uses Tentris' multi-way join algorithm (Algorithm \ref{alg:mwlj}, line 22), and tensor slicing operations to generate the operands of GraphQL queries.
\par
To bridge the gap between GraphQL schemata and RDF graphs, we follow Neo4j's example\footnote{\url{https://github.com/neo4j-graphql/neo4j-graphql-js/blob/master/docs/graphql-schema-directives.md}} and define several \emph{directives} in our implementation.
As per the GraphQL specification, ``directives can be used to describe additional information for types, fields, fragments and operations" \cite[Section 3.13]{url/graphql}.
We mentioned in Section \ref{sec:gqlrdf} that GraphQL types and fields need to be mapped to RDF terms.
To this end, we define in our implementation the directive \texttt{@uri}.
For example, the type definition \texttt{type Person @uri(value: "http://www.exmpl.org/Person")} maps the type \texttt{Person} to the RDF term \texttt{http://www.exmpl.org/Person}.
As the inverse of a property is not always available in RDF graphs, we also define the field directive \texttt{@inverse}, which denotes that the inverse direction of a field's property should be used.
Last, we introduce the field directive \texttt{@filter}, which denotes that the results of a particular field should be filtered using that field's type.
This directive should be used on fields that are mapped to properties having ranges consisting of multiple RDF classes (Section \ref{sec:gqlrdf}).

\section{Experimental Results}\label{sec:evaluation}
In this section, we present the performance evaluation of TentrisGQL, which we evaluated using the Link{\"o}ping GraphQL Benchmark (LinGBM) \cite{DBLP:conf/wise/ChengH22}.
LinGBM is a synthetic benchmark generator that provides a GraphQL schema that captures the structure of the generated datasets, and a set of 16 GraphQL query templates.
To the best of our knowledge, LinGBM is currently the only publicly available benchmark for evaluating GraphQL services.
The experiments that are presented below were carried out on a Debian 10 server with an AMD EPYC 7742 64-Core Processor, 1TB RAM, and two 3 TB NVMe SSDs in RAID 0.
All artifacts (e.g., datasets, GraphQL schemata, queries, and system configurations) are available online.\footnote{\url{https://github.com/dice-group/graphql-benchmark}}

\subsection{Systems}
As baseline for our experiments, we used Neo4j Community Edition 5.5.0 \cite{url/neo4j}.
We selected Neo4j because it is a widely used graph database and it provides its own tools for processing GraphQL queries.
In our experiments, we evaluated Neo4j in two different modes.
In the first mode (Neo4jC), Neo4j was provided with Cypher queries instead of GraphQL queries.
The GraphQL queries used in our experiments (Section \ref{sec:dqs}) were translated to Cypher queries using a library provided by Neo4j\footnote{\url{https://github.com/neo4j-graphql/neo4j-graphql-js}}.
The purpose of this mode was to compare the query evaluation performance of TentrisGQL against that of Neo4j, as no result rewriting takes place in this mode of Neo4j.
To find out the overhead introduced by the process of result rewriting, we used a second mode, namely Neo4jGQL.
Neo4jGQL includes an external application that is connected to Neo4j and is responsible for translating GraphQL queries to Cypher queries and rewriting query results to GraphQL responses.\footnote{We followed the example used in \url{https://github.com/neo4j-graphql/neo4j-graphql-js}.}
Recall that TentrisGQL incrementally constructs GraphQL responses.
For the evaluation, we used Neo4j's recommended memory settings\footnote{\url{https://neo4j.com/docs/operations-manual/5/tools/neo4j-admin/neo4j-admin-memrec}} and built the appropriate search indices.
More specifically, regarding the memory settings, we allocated 31GB of memory to the Java virtual machine (JVM) and 957GB for caching purposes.
In our experiments, we also evaluated TentrisGQLBase, a version of TentrisGQL that treats fields of type ID (i.e., fields that capture IRIs of RDF terms) as strings.
As a result, TentrisGQLBase needs to carry out left joins and joins to evaluate such fields when they appear as leaf fields or arguments in a query, respectively.
In contrast, TentrisGQL accesses the IRIs of RDF terms directly.
TentrisGQLBase provides us with insights on the impact that the evaluation of leaf fields has on the performance of our service.

\subsection{Datasets, Query Templates, and Schema}\label{sec:dqs}
LinGBM's dataset generator relies on the dataset generator of LUBM \cite{DBLP:journals/ws/GuoPH05} and allows for the generation of datasets of varying sizes via the use of a scale factor. To evaluate the performance of our approach on RDF graphs of different sizes, we generated three graphs (Table \ref{table:datasets}), namely \emph{LinGBM100}, \emph{LinGBM500}, and \emph{LinGBM1000}. 
For our experiments, we modified LinGBM's dataset generator to include the classes corresponding to the interface types of the schema, as both systems expect them to be stated in the input data.
\par
As previously mentioned, LinGBM also provides a set of query templates.
Their design follows a choke-point methodology \cite[Section 3.3]{DBLP:conf/wise/ChengH22}; each choke-point focuses on a particular workload or operation.
In this work, we are interested in join and left-join operations.
Hence, we focus on the choke-points \emph{Attribute Retrieval} (CP1) and \emph{Relationship Traversal} (CP2) of LinGBM.
There are only six query templates (QT1-QT6) that focus exclusively on CP1 and CP2.
To include additional queries in our evaluation, we modified the query templates QT7-QT14 by removing those features that are not related to CP1 and CP2 (e.g., ordering, filtering, and pagination). 
In addition, we had to remove input objects from the query templates QT11-QT14, as they are currently not supported by our implementation.
Ultimately, in our experiments, we used 11 query templates and two non-parameterized queries (Table \ref{tab:queries}).\footnote{After our modifications, QT8 and QT11 do not have any parameters, and QT13 and QT14 are identical.}
\par
The GraphQL schema provided by LinGBM is meant to be used by GraphQL services that do not generate GraphQL schemata automatically (e.g., relational schema to GraphQL schema).
For our experiments, we modified the provided schema by removing those types and features that are not required by the query templates (e.g., input types and enumeration types) used in the experiments. 
For TentrisGQL, we extended the schema with the directives of our implementation (Section \ref{sec:impl}).
In a similar manner, we extended the schema used for Neo4j with Neo4j's respective directives.
\begin{table}[t]
    \centering
    \begin{tabular}{@{}rccccc@{}}
        \toprule
        & Scale Factor & \#Triples & \thead{\#Distinct \\ Subjects} & \thead{\#Distinct \\ Predicates} & \thead{\#Distinct \\ Objects}  \\ \midrule
        LinGBM100  & 100  & 16M   & 2M     & 20  & 3M   \\
        LinGBM500  & 500  & 79M   & 10M    & 20  & 18M  \\
        LinGBM1000 & 1000 & 160M  & 21M    & 20  & 37M      \\
        \bottomrule
    \end{tabular}
    \caption{The datasets used in the experiments.}
    \label{table:datasets}
\end{table}
\begin{table}[t]
    \centering
    \begin{tabular}{ccccccccccc} \toprule
         QT & D & aRS-100 & aRS-500 & aRS-1000 & & QT & D & aRS-100 & aRS-500 & aRS-1000  \\ \midrule
         1  & 3 & 34K   & 170K & 338K   & & 8   &  1   & 4M   & 20M   & 40M  \\
         2  & 3 & 21K   & 105K & 200K   & & 9   &  4 & 279K & 1.3M  & 2.7M \\ 
         3  & 4 & 243   & 244  & 245    & & 10  & 1 & 6.3M & 31M   & 63M  \\ 
         4  & 5 & 81K   & 425K & 864K   & & 11  & 2 & 3.7M & 18M   & 37M \\ 
         5  & 7 & 12M   & 311M & 1.2G   & & 12  & 3 & 79K & 397K  & 785K \\ 
         6  & 4 & 19K   & 94K  & 192K   & & 13  & 3 & 73K & 373K  & 738K \\ 
         7  & 3 & 20K   & 101K & 202K   & & \\  
         \bottomrule
\end{tabular}
    \caption{The depth (D) and the average size of the GraphQL response (aRS-SF) in bytes of each GraphQL query template (QT) for each scale factor (SF). QT8 and QT11 are not parameterized.}
    \label{tab:queries}
\end{table}
\subsection{Benchmark Configurations and Execution}
Our GraphQL service was evaluated on two different benchmark configurations. 
The purpose of the first configuration was to evaluate the performance of our service on each query template.
For each template, we created a stress test consisting of ten query instances per template (110 queries)
We also created a stress test for each non-parameterized query (112 queries in total).
The stress tests were executed 5 consecutive times and independently from each other, thus ensuring that the query instances were executed the same number of times.
With the second configuration, we measured the performance of our system when queried by multiple clients.
This configuration consisted of one stress test, which included one query instance from each template and the two non-parameterized queries (i.e., 13 queries).
During the execution of the second configuration, it was important that there were multiple clients issuing queries at all times.
For this reason, we configured the clients to issue queries concurrently for one hour \cite{DBLP:conf/semweb/BigerlCBSSN20, DBLP:conf/semweb/ConradsLSMN17}.
In both configurations, the execution of each stress test is preceded by a warm-up run, in which the queries of the corresponding stress test are executed once.
This allowed Neo4j to load and cache its data structures in the main memory.
\par
The stress tests of both benchmark configurations were executed over HTTP using the benchmark execution framework IGUANA in version 3.3.0 \cite{DBLP:conf/semweb/ConradsLSMN17}.
As in \cite{DBLP:conf/semweb/BigerlCBSSN20}, we set the timeout across all benchmarks to three minutes and measured the performance of our implementation using the number of queries executed per second (QPS) and the penalized average QPS (pAvgQPS);
the penalty for failed queries (e.g., timed out queries) was set to three minutes.
Last, we compared the results generated by all systems to ensure that they return the same results across all queries.
\begin{figure}[t]
    \subfloat{%
        \includegraphics[width=0.32\textwidth]{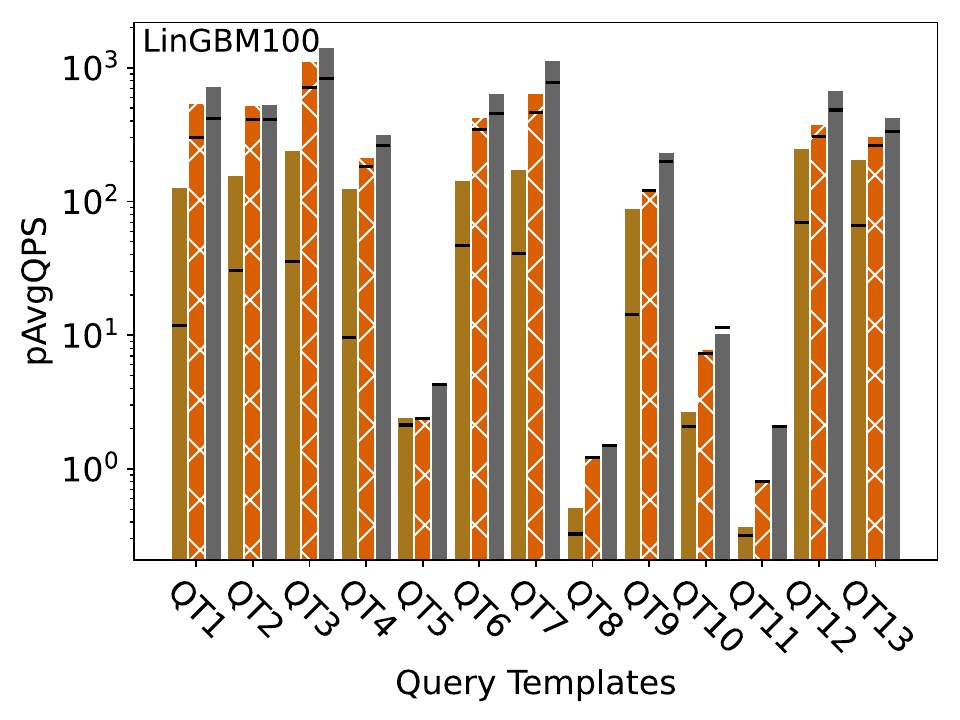}
    }
    \subfloat{%
        \includegraphics[width=0.32\textwidth]{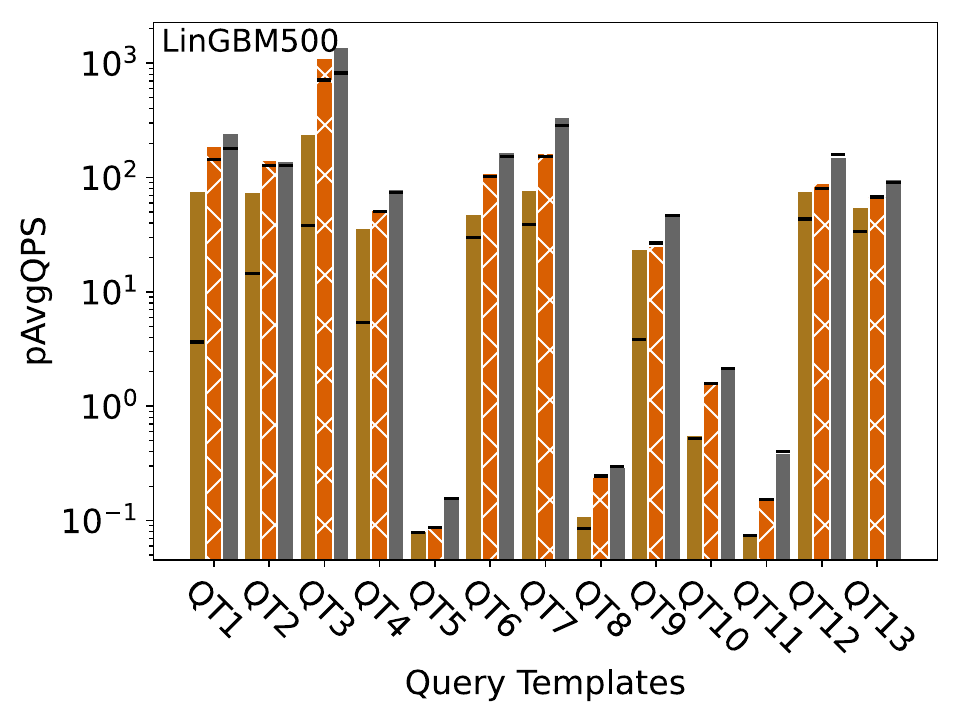}
    }
    \subfloat{%
        \includegraphics[width=0.32\textwidth]{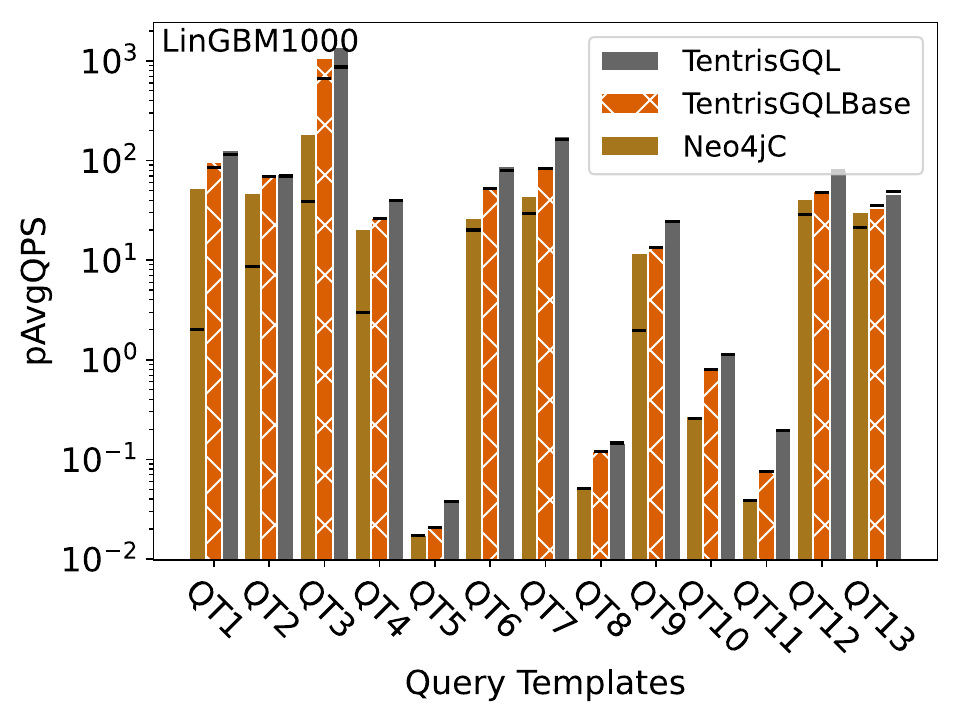}
    }
    \caption{Performance of the systems in the first configuration w.r.t. their pAvgQPS. The black lines denote the values reported in the warmup run.}
    \label{fig:avgqps}
\end{figure}
\begin{figure}[t]
    \subfloat{%
        \includegraphics[width=0.32\textwidth]{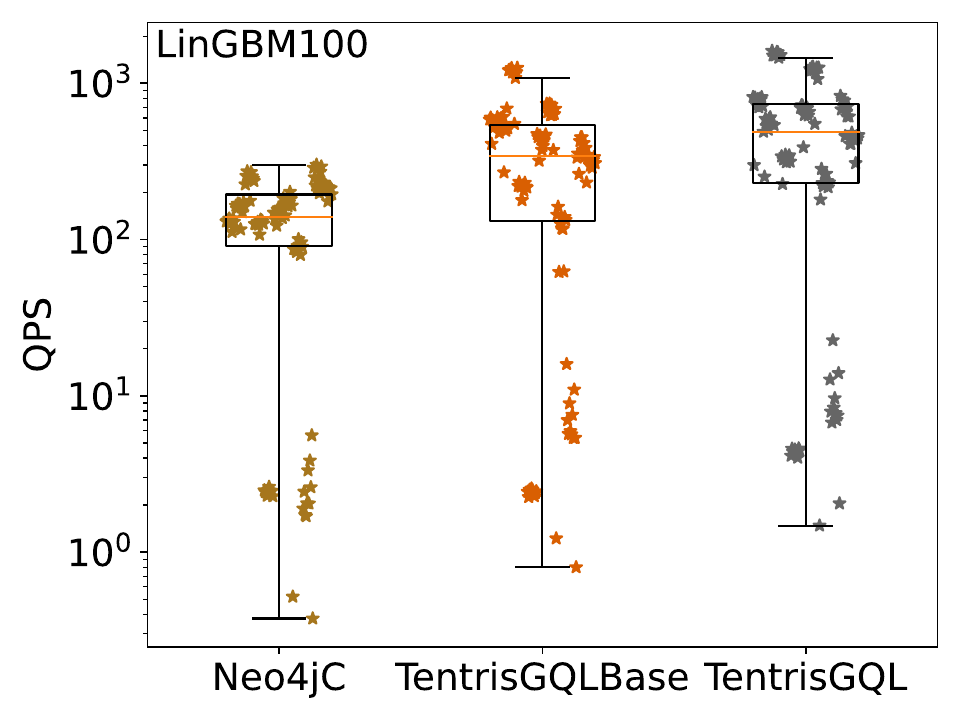}
    }
    \subfloat{%
        \includegraphics[width=0.32\textwidth]{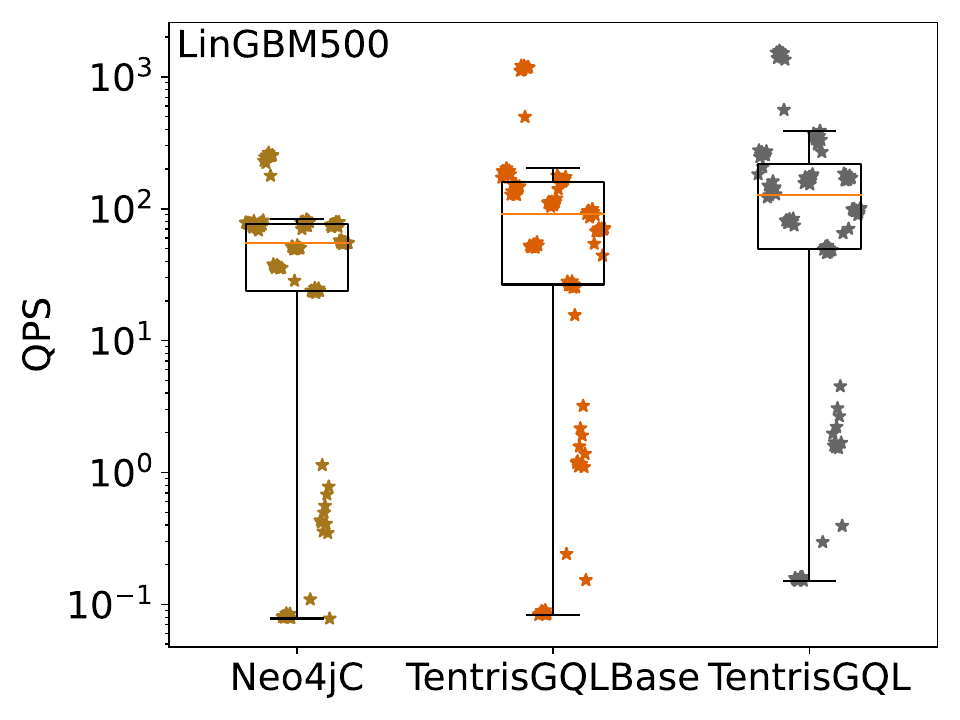}
    }
    \subfloat{%
        \includegraphics[width=0.32\textwidth]{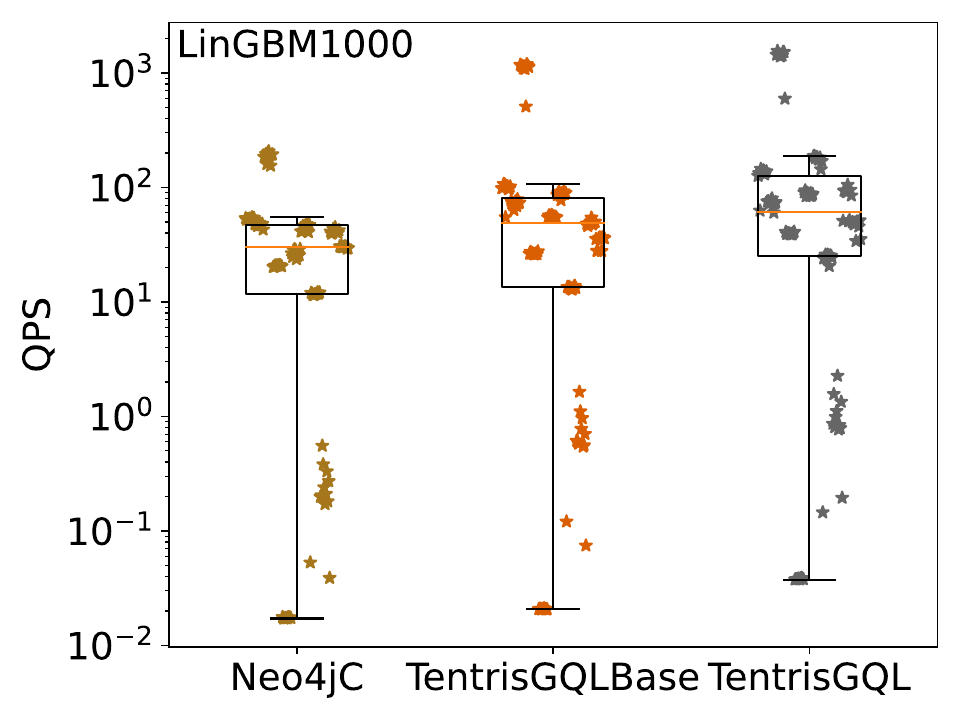}
    }
    
    \caption{Performance of the systems in the first configuration w.r.t. their QPS.}
    \label{fig:qps}
\end{figure}
\begin{figure}[t]
    \subfloat{%
        \includegraphics[width=0.33\textwidth]{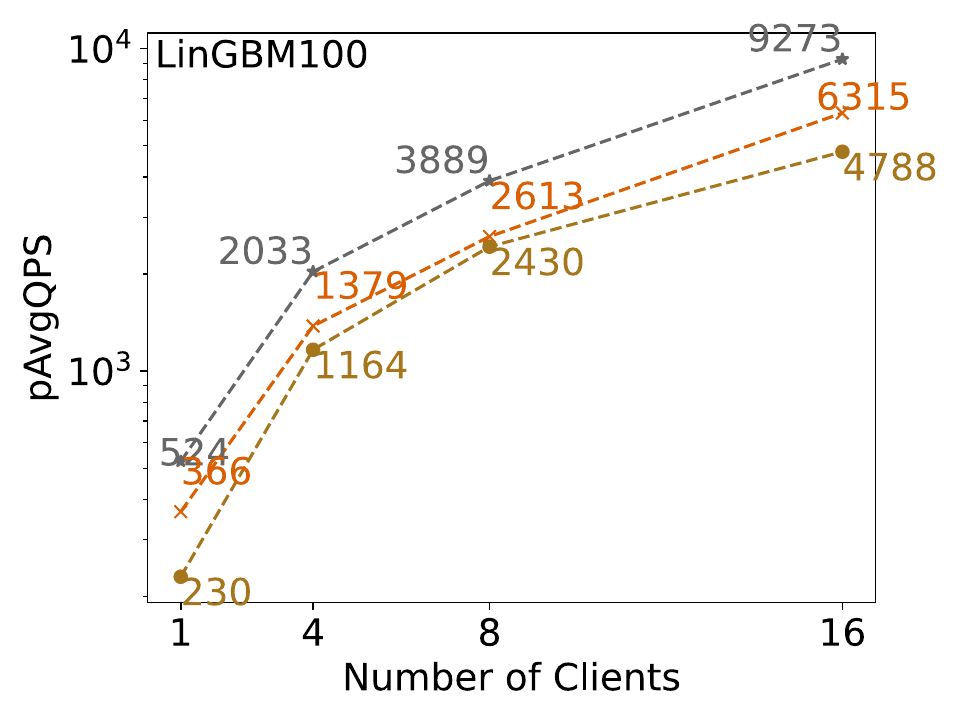}
    }
    \subfloat{%
        \includegraphics[width=0.33\textwidth]{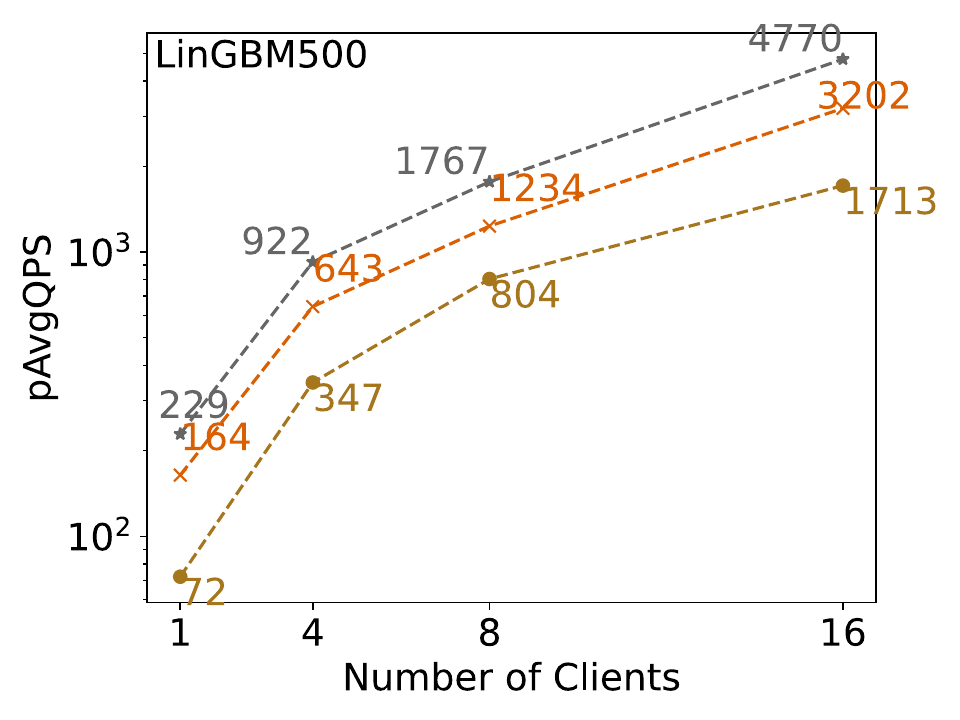}
    }
    \subfloat{%
        \includegraphics[width=0.33\textwidth]{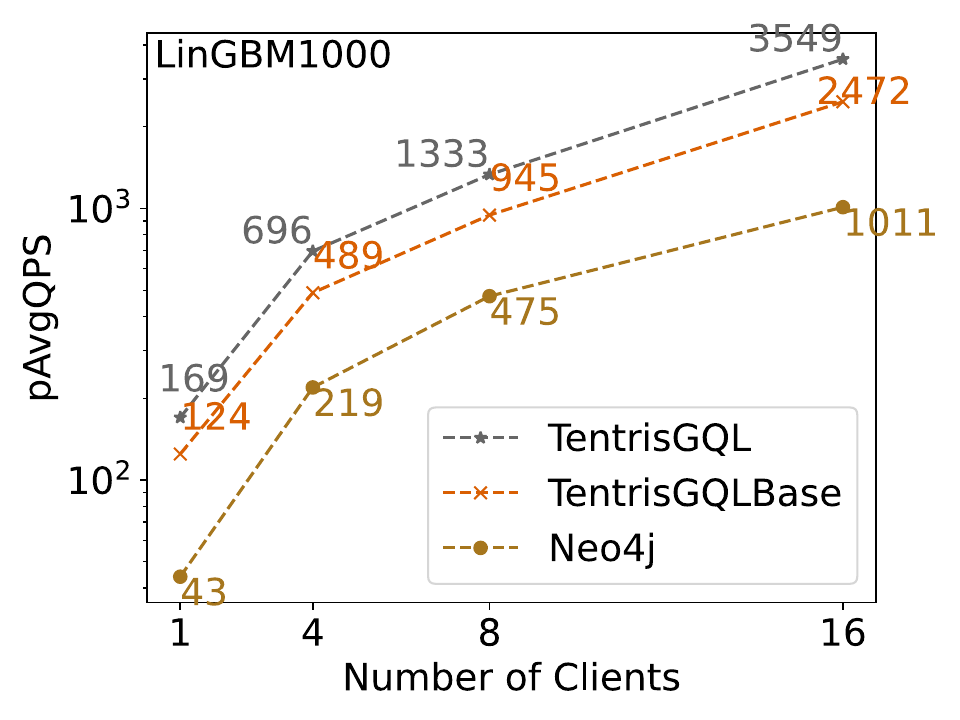}
    }
    \caption{Scalability of the systems in the second configuration w.r.t. their pAvgQPS.}
    \label{fig:scalability}
\end{figure}
\begin{table}[t]
    \centering
    \begin{tabular}{@{}rcccc@{}}
        \toprule
                   & Neo4jC  & Neo4jGQL  & TentrisGQLBase & TentrisGQL \\ \midrule
        LinGBM100  & 230.32  & 91.51     & 366.27         & 524.90     \\
        LinGBM500  & 72.13   & 33.72     & 164.23         & 229.90     \\
        LinGBM1000 & 43.99   & 20.11     & 124.59         & 169.79     \\
        \bottomrule
    \end{tabular}
    \caption{Overhead of result rewriting in Neo4jGQL (pAvgQPS).}
    \label{tab:rewrting}
\end{table}
\subsection{Results}
The results of the first benchmark configuration are presented in Figures \ref{fig:avgqps} and \ref{fig:qps}.
Figure \ref{fig:avgqps} shows that both TentrisGQL and TentrisGQLBase outperform Neo4jC across all query templates in all datasets, with TentrisGQL achieving 1.5 (QT2 and QT13) to 7.4 (QT3) times higher pAvgQPS than Neo4jC in the largest dataset, namely LinGBM1000.
In addition, both TentrisGQL and TentrisGQLBase achieve higher median QPS  than Neo4jC in all datasets (Figure \ref{fig:qps}).
Figure \ref{fig:scalability} summarizes the results reported in the second benchmark configuration.
We removed the query instance corresponding to QT5 from the second configuration's query list 
because Neo4j was running out of JVM memory when this query was issued by multiple clients.
TentrisGQL and TentrisGQLBase did not face any memory-related issues.
Figure \ref{fig:scalability} shows that Neo4jC scales better than TentrisGQL and TentrisGQLBase when queried by 4 and 8 concurrent clients in the smallest dataset (i.e., LinGBM100).
However, TentrisGQL and TentrisGQLBase achieve higher pAvgQPS than Neo4jC in all cases and in particular, TentrisGQL achieves 3.5 higher pAvgQPS in the case of 16 clients in LinGBM1000.
To measure the overhead introduced by the rewriting of Neo4j's results to GraphQL responses, we used the second benchmark configuration with one concurrent user.
Table \ref{tab:rewrting} shows that rewriting process leads to TentrisGQL achieving up to 8.4 times higher pAvgQPS than Neo4jGQL 
\subsection{Discussion}
The performance of the systems did not vary significantly across all datasets. In particular, they were not significantly affected by the increasing size of the datasets.
The systems' performance was mostly affected by the average result size (aRS) of the query templates (Table \ref{tab:queries}).
In particular, all systems achieved their highest and lowest pAvgQPS in all datasets in QT3 and QT5, respectively. 
QT3 has the lowest aRS, whereas QT5 has the highest.
The depth of the query templates also affects the systems' performance.
For example, the pAvgQPS of the systems in QT6 is lower than in QT7, even though the latter has a higher aRS.
Queries with higher values of depth require more left-join operations in TentrisGQL and longer path traversals in Neo4j.
\par
Another factor that impacts the performance of our GraphQL service is the size of the operands corresponding to leaf fields, which are evaluated via left-join operations.
This observation is grounded in the performance of TentrisGQLBase (Figure \ref{fig:avgqps}), which is always equivalent to, or worse than, TentrisGQL's performance.
Recall that TentrisGQLBase, unlike TentrisGQL, evaluates leaf fields and arguments of type ID via left-join and join operations, respectively.
Neo4j employs the property graph model, which allows it to represent leaf fields as node properties.
Hence, for evaluating leaf fields, Neo4j does not iterate over all of a particular property key's properties.
Despite these additional operations, both TentrisGQL and TentrisGQLBase outperform Neo4j.
This suggests that our algorithm does not introduce much overhead to the computations.
\par
The results of Table \ref{tab:rewrting} are in line with the results reported in \cite{DBLP:conf/semweb/GleimHKD20} and demonstrate the importance of GraphQL services being able to directly construct GraphQL responses.
Regarding the memory usage, we measured the memory used by the systems in LinGBM1000 when queried by 16 concurrent clients using \texttt{pmap}\footnote{\url{https://linux.die.net/man/1/pmap}}.
The highest Resident Set Size (RSS) reported by TentrisGQL and Neo4jC was 41GB and 44GB, respectively.
\section{Related Work}\label{sec:relwork}
Recently, several graph storage solutions have made efforts to allow users to access their data via GraphQL.
Dgraph \cite{url/dgraph} is a distributed graph database that natively supports GraphQL. It also provides its own query language, namely DQL.
In Dgraph, GraphQL operations are translated to DQL operations.
However, response objects are constructed following the GraphQL specification.
Hence, a rewriting of the results is not required. We did not include Dgraph in our experiments for two reasons.
First, Dgraph does not fully support RDF, as it is not able to handle URIs.
Additionally, Dgraph's GraphQL service expects predicates to be prefixed with their subject's type.
Consequently, existing RDF graphs need to be substantially modified to be stored in a Dgraph instance.
Second, Dgraph does not provide a bulk loader for its GraphQL service; hence it is not able to load large knowledge graphs efficiently.\footnote{\url{https://discuss.dgraph.io/t/graphql-vs-dql-dgraph-blog/14311}; see paragraph "When not to use GraphQL".}
In addition to the translation tools used in Section \ref{sec:evaluation}, Neo4j provides a library that serves as a middleware between applications and database instances.
This library\footnote{\url{https://neo4j.com/docs/graphql-manual/current/}} is responsible for the translation process of GraphQL queries to Cypher queries.
Regarding triple stores, Stardog \cite{url/stardog} and the commercial edition of GraphDB \cite{url/graphdb} provide GraphQL support by translating GraphQL to SPARQL.
Virtuoso \cite{url/virtuoso} introduced a GraphQL plugin\footnote{\url{https://community.openlinksw.com/t/introducing-native-graphql-support-in-virtuoso/3378}} that allows its users to query RDF graphs via GraphQL.
To bridge the gap between GraphQL and SPARQL, this plugin relies on OWL ontologies to map the types and fields of GraphQL schemata to RDF terms.
We did not include Virtuoso in our experiments as it does not perform type filtering in inner fields, which leads to queries returning incorrect results.\footnote{\url{https://github.com/openlink/virtuoso-opensource/issues/1115}}

\section{Conclusion and Future Work}\label{sec:conclusion}
We presented an approach for the native evaluation of GraphQL queries over RDF graphs. 
As GraphQL queries require left-join operations, we focused on the development of a novel multi-way left-join algorithm that is inspired by worst-case optimal multi-way join algorithms.
Similarly to worst-case optimal multi-way join algorithms, the proposed left-join algorithm recursively evaluates queries on a per variable basis, which allows for the incremental enumeration of GraphQL queries.
By implementing our approach within the tensor-based triple store Tentris, we provide the first publicly available triple store that treats GraphQL as a first-class citizen.
The performance evaluation of our implementation demonstrates the efficiency of the left-join algorithm, as our implementation outperforms a state-of-the-art graph database, namely Neo4j.
\par 
Our implementation currently supports the features of the language that are required by its formal semantics (Equation \ref{eq:g_semantics}).
Our future work will focus on extending our GraphQL service with all features from the specification.
To the best of our knowledge, there have not been any works that focus on evaluating SPARQL queries requiring left-join operations on a variable basis.
To this end, we plan to use our approach for the evaluation of such SPARQL queries (i.e., queries containing optional graph patterns).

\bibliographystyle{vancouver}
\bibliography{bibliography}

\end{document}